\documentstyle[psfig, referee]{mn}
\topmargin -0.5cm \title[]{Light curves of jetted gamma-ray burst afterglows 
                                                in circumstellar clouds}  
 
\author[]{Z. G. Dai, Y. F. Huang, and T. Lu\\
\rm Department of Astronomy, Nanjing University, Nanjing 210093, China}
  
\date{(accepted for publication in {\em MNRAS Letters})}
\pubyear{2001}

\begin{document}

\maketitle 

\begin{abstract}
The afterglow emission from a  spreading jet expanding in a circumstellar cloud 
is discussed. Prompt X-ray radiation and a strong UV flash from the reverse shock 
produced by the interaction of the jet with the cloud may destroy and clear 
the dust out to about 30 pc within the initial solid angle of the jet. As the 
sideways expansion of the jet becomes significant, most of the optical radiation 
from the high-latitude part of the jet may be absorbed by the dust outside 
the initial solid angle of the jet, but only the radiation from the part 
within the initial solid angle can be observed. We analytically show 
that the flux of the observational radiation decays as $\propto t^{-(p+1)}$ 
(where $p$ is the power-law index of the electron distribution) in the 
relativistic phase. This preliminary result motivates us to perform  
numerical calculations. Our results show that one break in the 
optical afterglow ligh curve extends over a factor of $\sim 3$ in time 
rather than one decade in time in the previous jet model. 
These results may provide a way to judge whether GRBs locate
in dense clouds or not. Finally, we carry out a detailed modelling 
for the R-band afterglow of GRB 000926.    
\end{abstract}

 
\section {Introduction}

The origin of gamma-ray bursts (GRBs) has puzzled us for more than three 
decades (see, e.g., Wijers, Rees \& M\'{e}sz\'{a}ros 1997; Wijers 1998). 
In the study of GRBs, two fundamental questions 
pertinent directly to the central engine need to be solved 
urgently and unambiguously. One question is about the existence of jets 
in GRBs. Multi-wavelength observations of afterglows have provided several 
indications of jetted GRBs: theoretically, it is first shown analytically that 
the sideways expansion (Rhoads 1999; Sari, Piran \& Halpern 1999) and edge 
effects of jets (M\'esz\'aros \& Rees 1999) can lead to a marked break in afterglow 
light curves, and subsequent calculations indicate that  one break appears but
extends over at least one decade in time (Moderski, Sikora \& Bulik 2000;
Kumar \& Panaitescu 2000; Huang et al. 2000a, b, c; Wei \& Lu 2000). 
These effects seem to account for the observed light curves 
of the optical afterglows from GRB 990123 (Kulkarni et al. 1999; 
Castro-Tirado et al. 1999; Fruchter et al. 1999), GRB 990510 
(Harrison et al. 1999; Stanek et al. 1999), GRB 990705 (Masetti et al. 2000a), 
GRB 991208 (Sagar et al. 2000a; Galama et al. 2000a), GRB 991216 (Halpern 
et al. 2000a), and GRB 000301C (Rhoads \& Fruchter 2001; Masetti et al. 2000b; 
Jensen et al. 2000; Berger et al. 2000; Sagar et al. 2000b). Further evidence for 
a jet-like geometry in GRBs is the observed radio flares from GRB 990123 
(Kulkarni et al. 1999) and GRB 990510 (Harrison et al. 1999). In addition, 
the polarization observed in some afterglows also provides an important 
signature for jetted GRBs (Wijers et al. 1999).  
 
Another question is about the association of GRBs with star-forming regions. 
There is considerable evidence linking the progenitors of GRBs with massive stars.   
For example,  the sources of the GRBs with known redshifts lie within the optical 
radii and central regions of the host galaxies rather than far outside the disks of 
the galaxies (Bloom, Kulkarni \& Djorgovski 2000), which seems to rule out mergers 
of neutron-star binaries as the GRB central engine. Further evidence has been provided 
by the fact that the brightness distribution of GRBs is in agreement with the models 
in which the GRB rate tracks the star formation rate over the past 15 billion years of 
cosmic history (Totani 1997; Wijers et al. 1998; Kommers et al. 2000). The most 
direct evidence for the relation between GRBs and a specific type of supernova 
(i.e., hypernovae/collapsars) is the discovery of SN 1998bw in the error box of 
GRB 980425 (Galama et al. 1998) and the detection of a supernova-like 
component in the afterglows from GRB 980326 (Bloom et al. 1999) and 
GRB 970228 (Reichart 1999; Galama et al. 2000b). Finally, the recent discovery 
of a transient absorption edge in the X-ray spectrum of GRB 990705 
(Amati et al. 2000) and the observations of X-ray lines from GRB 991216 
(Piro et al. 2000) and GRB 000214 (Antonelli et al. 2000) provide new evidence 
that GRBs are related to the core collapse of massive stars.  Based on these 
observational facts, it is natural to suppose that the environments of GRBs should be 
pre-burst winds and/or circumstellar clouds. If GRBs occur in such  
pre-burst winds, their afterglows will fade down rapidly (Dai \& Lu 1998;
Chevalier \& Li 1999, 2000); if GRBs are surrounded by the circumstellar clouds, 
their fireballs (or jets) must evolve to the non-relativistic regime within a few 
days after the bursts, leading to a steepening of the afterglow light curves 
(Dai \& Lu 1999, 2000; Wang, Dai \& Lu 2000).       

The purpose of this paper is to combine these two fundamental points to 
discuss their implications instead of directly proving their existence. We assume 
that a GRB comes from a highly collimated jet expanding in a circumstellar 
cloud, and discuss  the resultant afterglow emission when the jet is spreading 
laterally. In section 2, we analyze the motivation of our work: a strong 
UV flash from the reverse shock produced by the interaction of the jet with 
the cloud may clear the dust out to about 30 pc only along the initial path of 
the GRB. As a result, most of the optical radiation from the high-latitude part of 
the jet may be strongly extincted by the dust outside the initial solid angle 
of the jet, but only the radiation from the part within the initial solid angle can 
be observed. One expects that this could produce a rapidly fading afterglow. 
In section 3, we further analyze the spectrum and light curve. In section 4, 
we present our numerical procedure and results. We carry out 
a detailed modelling for the R-band afterglow of GRB 000926 in section 5, 
and give a brief discussion and summary in the final section.  

\section{Motivation}

In the standard afterglow model (for recent reviews see Piran [1999] and 
van Paradijs, Kouveliotou \& Wijers [2000]), a GRB relativistic 
shell with a Lorentz factor of $\eta$ is assumed to interact with the ambient medium
(circumstellar cloud) via two shocks: a reverse shock and a forward shock. The 
forward shock runs forward into the cloud while the reverse shock sweeps up the 
shell material. The observed prompt optical flash of GRB 990123 has 
been argued to come from a reverse shock (Akerlof et al. 1999; Sari \& Piran 1999; 
M\'esz\'aros \& Rees 1999). We believe that reverse shock emission should be 
common for all GRBs. The shocked cloud and shell materials are in pressure 
balance and are separated by a contact discontinuity. Since an optical flash is 
produced when the reverse shock crosses the shell, the shocked materials 
expand with a Lorentz factor of $\gamma$, which is approximated by the 
Blandford-MeKee's self-similar solution. As argued by Waxman \& Draine 
(2000), the reverse shock (with a Lorentz factor of $\gamma^{\rm rs}$) may 
be mildly relativistic, i.e., $|\gamma^{\rm rs}-1|\sim 1$, and thus $\gamma
\sim \eta$. We assume that the power-law index of the electron distribution, 
$p$, is similar in the forward and reverse shocks. If the fractions of the 
thermal energy carried by electrons, $\epsilon_e$, and magnetic field, 
$\epsilon_B$, are also similar in the two shocks, then the typical electron Lorentz 
factor in the reverse shock $\gamma^{\rm rs}_m\approx 610\xi\epsilon_e$ 
and the magnetic field $B\approx (32\pi \eta^2nm_pc^2)^{1/2}$, where 
$\xi=3(p-2)/(p-1)\approx 1$ and $n$ is the baryon number density of the cloud. 
 
We consider synchrotron radiation from the reverse shock. To calculate the 
luminosity of the optical flash, one needs to know the emission 
spectrum, which is determined by two break frequencies: the peak frequency 
$\nu^{\rm rs}_m$ and the cooling frequency $\nu^{\rm rs}_c$.  We first derive 
the observed peak synchrotron frequency: 
\begin{equation}
\nu^{\rm rs}_m\approx 5.8\times 10^{12}\xi^2\epsilon_{e,-1}^2
\epsilon_{B,-6}^{1/2}\eta_{300}^2n_3^{1/2}\left(\frac{1+z}{2}\right)^{-1}\,{\rm Hz}, 
\end{equation}
where $\epsilon_{e,-1}=\epsilon_e/0.1$, $\epsilon_{B,-6}=
\epsilon_B/10^{-6}$, $\eta_{300}=\eta/300$, and $n_3=n/10^3\,{\rm cm}^{-3}$. 
The cooling frequency can be approximately expressed as 
\begin{equation}
\nu^{\rm rs}_c\sim 10^{19}\epsilon_{B,-6}^{-3/2}\varepsilon_{54}^{-1/2}n_3^{-1}
t_1^{-1/2}\left(\frac{1+z}{2}\right)^{1/2}\,{\rm Hz}, 
\end{equation}
where $t_1$ is the observer's time ($t$) in units of 10 s  
(Wang, Dai \& Lu 2001). In addition, using equation (4) of Waxman \& Draine 
(2000), we find the observed peak synchrotron flux density for the reverse shock 
\begin{equation}
F^{\rm rs}_{\nu_m}\approx 40\epsilon_{B,-6}^{1/2}\eta_{300}^{-1}n_3^{1/4}
\varepsilon_{54}^{5/4}t_1^{-3/4}\left(\frac{1+z}{2}\right)^{3/4}
\left(\frac{\sqrt{1+z}-1}{\sqrt{2}-1}\right)^{-2}\,{\rm Jy}, 
\end{equation}
where $\varepsilon=\varepsilon_{54}\times 10^{54}\,{\rm ergs}\,\,{\rm sr}
^{-1}$ is the shell energy per unit solid angle, $z$ is the redshift of the source, and 
a flat universe with zero cosmological constant and the Hubble constant $H_0=65\,
{\rm km}\,{\rm s}^{-1}\,{\rm Mpc}$ is assumed. Therefore, we get the spectrum 
\begin{equation}
F^{\rm rs}_\nu=\left \{
       \begin{array}{ll}
           (\nu/\nu^{\rm rs}_m)^{-(p-1)/2}F^{\rm rs}_{\nu_m} & {\rm if}\,\, 
              \nu^{\rm rs}_m<\nu<\nu^{\rm rs}_c, \\ 
           (\nu^{\rm rs}_c/\nu^{\rm rs}_m)^{-(p-1)/2}(\nu/\nu^{\rm rs}_c)^{-p/2}
              F^{\rm rs}_{\nu_m} & {\rm if}\,\, \nu>\nu^{\rm rs}_c. 
       \end{array}
       \right.
\end{equation}
From this spectrum, we can easily calculate the  
local-frame prompt luminosity ($L_{1-7.5}$) in the $1-7.5$ eV range 
({\em UV band}). For example, if $p=2.5$, we find 
\begin{equation}
L_{1-7.5}\sim 1\times 10^{50}\epsilon_{e,-1}^{3/2}\epsilon_{B,-6}^{7/8}
\eta_{300}^{1/2}n_3^{5/8}\varepsilon_{54}^{5/4}t_1^{-3/4}
\left(\frac{1+z}{2}\right)^{3/4}\,{\rm ergs}\,\,{\rm s}^{-1}.
\end{equation} 
Please note that Waxman \& Draine (2000) calculated the UV luminosity 
at $p=2$. Here we have extended their discussion to any given $p$. 

We turn to discuss dust destruction in a circumstellar cloud with a typical radius 
of $R_c\sim 30\,$pc. Since the column density of the cloud is $N_H=nR_c\approx 
10^{23}n_3(R_c/30\,{\rm pc})\,\,{\rm cm}^{-2}$, the inferred visual 
extinction is $\sim 50n_3(R_c/30\,{\rm pc})$ if the cloud  is assumed to have 
the same dust-to-gas ratio as the Galactic clouds, implying that an afterglow from 
the GRB would be completely extincted without dust destruction. Fortunately, 
as argued by Waxman \& Draine (2000), a strong UV flash from the GRB can 
destroy its ambient dust by thermal sublimation. Using their equation (17), we 
find that the dust grains are completely sublimed out to a destruction radius 
$R_d\approx 30(L_{1-7.5}/10^{50}{\rm ergs}\,\,{\rm s}^{-1})^{1/2}$ pc, 
where the grain radius $a=0.1\,\,\mu$m is assumed. Thus,   
a strong UV flash with luminosity of $\sim 10^{50}\,\,{\rm ergs}\,\,{\rm s}^{-1}$, 
if isotropic, can clear all the dust grains in a typical giant molecular cloud.
Furthermore, Fruchter, Krolik \& Rhoads (2001) show that if a GRB emits X-rays
in a way similar to those observed by BeppoSAX, dust grains along the line 
of sight at a distance as large as $\sim 100$ pc in the host galaxy of the 
burst can be destroyed by these prompt X-rays due to grain heating and charging. 
These results were recently confirmed by Galama \& Wijers (2001), who analyzed 
a complete sample of GRB afterglows, and found that the GRB environments 
have both high column densities ($\sim 10^{23}\,\,{\rm cm}^{-2}$) of gas and 
low optical extinctions. However, since the GRB is believed to come from 
a jet with the initial half opening angle of $\theta_0\sim 0.1$, the UV flash 
and prompt X-ray emission will clear the dust only 
along the initial path of the burst. Therefore, the optical 
radiation from the $\theta\le\theta_0$ part has low extinction but the optical 
radiation from the high-latitude ($\theta>\theta_0$) part could be absorbed 
by the dust outside the initial solid angle of the jet, implying that only 
an afterglow emission from the low-latitude of the jet would be observed 
if the line of sight is along the jet axis. This could lead to a rapidly fading
optical afterglow. Motivated by this argument, we will discuss the afterglow 
emission from a  spreading jet expanding in a circumstellar cloud
in the next section. 

\section{Analytical Model: Spectrum and Light Curve}

Rhoads (1999) has considered the evolution of a {\em relativistic} jet (with 
Lorentz factor of $\gamma$) that is spreading laterally at the local speed of 
sound $c_s=c/\sqrt{3}$ (but $c_s=c$ in Sari et al. [1999]), so the half opening 
angle $\theta_j\sim \theta_0+\gamma^{-1}/\sqrt{3}$. In this case, the dynamical 
transition takes place at $\gamma\sim \theta_0^{-1}/\sqrt{3}$. In the initial stage 
of the evolution, since $\gamma>\theta_0^{-1}/\sqrt{3}$, the jet is spherical-like 
and its Lorentz factor decays as $\gamma\propto t^{-3/8}$.  The resulting spectrum 
and light curve are well known (Sari, Piran \& Narayan 1998): 
\begin{equation}
F^{\rm fs}_\nu\propto\left \{
       \begin{array}{ll}
          \nu^{-(p-1)/2}t^{-3(p-1)/4} & {\rm if } \,\, \nu^{\rm fs}_m<\nu<\nu^{\rm fs}_c,\\
          \nu^{-p/2}t^{-(3p-2)/4} & {\rm if} \,\, \nu>\nu^{\rm fs}_c.  
      \end{array}
      \right.
\end{equation} 
We next stress to discuss the afterglow emission for $\gamma\ll \theta_0^{-1}/\sqrt{3}$. 
In this spreading stage, the sideways expansion leads to an exponential decay 
of $\gamma$ with radius (Rhoads 1999; Sari et al. 1999). As a result, the jet's radius 
$R\propto t^0$ and $\gamma\propto t^{-1/2}$. The typical synchrotron frequency 
decays as $\nu^{\rm fs}_m\propto t^{-2}$ and the cooling frequency $\nu^{\rm fs}_c$ 
is a constant. From the discussion in section 2, the total number of 
electrons radiating toward to the observer, those located in a cone of 
the initial opening angle, can be estimated as $N_e=\pi\theta_0^2R^3n/3$. 
The total specific luminosity emitted by these electrons, $(1+z)\sigma_T
m_ec^2N_eB'\gamma/(3e)$, is distributed over an area of $\pi\gamma^{-2}
D_L^2$ at the luminosity distance $D_L$ from the source (where $B'$ is 
the magnetic field strength of the shocked cloud). Thus, the observed peak 
flux density is given by $F^{\rm fs}_{\nu_m}=(1+z)\sigma_Tm_ec^2R^3n
\theta_0^2B'\gamma^3/(9eD_L^2)\propto R^3\gamma^4\propto t^{-2}$. 
Therefore, we obtain the afterglow's spectrum and light curve:
\begin{equation}
F^{\rm fs}_\nu=\left \{
       \begin{array}{lll}
         (\nu/\nu^{\rm fs}_m)^{1/3}F^{\rm fs}_{\nu_m}\propto \nu^{1/3}t^{-4/3} &
             {\rm if}\,\, \nu<\nu^{\rm fs}_m, \\
         (\nu/\nu^{\rm fs}_m)^{-(p-1)/2}F^{\rm fs}_{\nu_m}\propto \nu^{-(p-1)/2}
              t^{-(p+1)} & {\rm if}\,\, \nu^{\rm fs}_m<\nu<\nu^{\rm fs}_c, \\
         (\nu^{\rm fs}_c/\nu^{\rm fs}_m)^{-(p-1)/2}(\nu/\nu^{\rm fs}_c)^{-p/2}
              F^{\rm fs}_{\nu_m}\propto\nu^{-p/2}t^{-(p+1)} & {\rm if}\,\, 
              \nu>\nu^{\rm fs}_c.
        \end{array}
       \right.
\end{equation}
It is easy to see that the temporal decay index of the high-frequency afterglow 
changes from $\alpha_1=3(p-1)/4$ or $(3p-2)/4$ to $\alpha_2=p+1$ because of 
the effects of sideways expansion and dust extinction. In conclusion, a marked 
break should appear in the light curve of the afterglow from a spreading jet 
expanding in the circumstellar giant cloud.  

\section{Numerical Results}

To prove the analytical result in section 3, we perform detailed numerical 
calculations for the evolution of the afterglow emission. We use the model proposed 
by Huang, Dai \& Lu (1999) and the calculational code developed by Huang et al. 
(2000c). This model has several advantages: (i) It is applicable to both radiative 
and adiabatic jets, and proper for both ultra-relativistic and non-relativistic 
stages. The model even allows the radiative efficiency $\epsilon$ to 
change with time, so that it can trace the evolution of a partially radiative jet
(Dai, Huang \& Lu 1999). However, note that some authors (e.g., M\'esz\'aros, Rees 
\& Wijers 1998) have argued that the jet should be adiabatic most likely, since it is 
unlikely that the radiative efficiency could reach 1; 
(ii) The model considers the lateral expansion of the jet. The evolution of 
the lateral speed (taken as the speed of sound) is given by a reasonable 
expression (see equation [12]). (iii) The model also takes into account many 
other effects, e.g., {\em the cooling of electrons and the equal arrival time surfaces}.  
Let $M_{\rm ej}$ be the initial ejecta mass and $m$ the swept-up cloud mass. 
The dynamical evolution of the jet 
is described by (Huang et al. 2000c)
\begin{equation}
\frac{d R}{d t} = \beta c \gamma (\gamma + \sqrt{\gamma^2 - 1}),
\end{equation}
\begin{equation}
\frac{d \theta_j}{d t} = \frac{c_s (\gamma + \sqrt{\gamma^2 - 1})}{R},
\end{equation}
\begin{equation}
\frac{d m}{d R} = 2 \pi R^2 (1 - \cos \theta_j) n m_{\rm p},
\end{equation}
\begin{equation}
\frac{d \gamma}{d m} = - \frac{\gamma^2 - 1}
       {M_{\rm ej} + \epsilon m + 2 ( 1 - \epsilon) \gamma m}, 
\end{equation}
\begin{equation}
c_s^2 = \frac{\hat{\gamma} (\hat{\gamma} - 1) (\gamma - 1)c^2}
{1 + \hat{\gamma}(\gamma - 1)}, 
\end{equation}
where $\beta=\sqrt{1-1/\gamma^2}$ and $\hat{\gamma}\approx (4\gamma+1)
/(3\gamma)$ (Dai et al. 1999). If $\epsilon \rightarrow 0$ {\em as in the following 
calculations}, equation (11) turns out to express the conservation of energy: 
$(\gamma-1)M_{\rm ej}+(\gamma^2-1)m={\rm constant}$ (for a discussion 
see Huang et al. [1999] and van Paradijs et al. [2000]).   

Figure 1 presents light curves of R-band afterglow emission ($p=2.5$) for two 
cases with dust extinction ({\em solid line}) and without dust extinction ({\em dashed 
line}). This figure clearly shows that the light curve with dust extinction is indeed 
more steepening than that without dust extinction at late times. Figure 2 
exhibits  $\alpha\equiv -d\ln F_R/d\ln t$ as a function of time for these two cases. 
For the case without dust extinction, $\alpha$ increases from $\sim 1.3$ at initial 
one day to $\sim 2.5$  at later times. This change extends over one order of 
magnitude in time, which is consistent with the previous numerical results 
(Moderski et al. 2000; Kumar \& Panaitescu 2000; Huang et al. 2000a, b, c). 
However, for the case with dust extinction,  $\alpha$ increases from $\sim 1.3$ 
at initial one day to $\sim 3.3$  at later times, which is in approximate accord
with the analytical result in section 3. Furthermore, the steepening is completed over 
a factor of $\sim 3$ in time, leading therefore to a sharper break in the light curve. 
 
\section{Afterglow of GRB 000926}

GRB 000926 was detected on 2000 September 26.9927 UT by the 
Inter-Planetary-Network (IPN) group of spacecrafts Ulysses, Russian Gamma-Ray
Burst Experimant (KONUS) and Near Earth Asteroid Rendezvous (NEAR)
(Hurley et al. 2000). The burst lasted $\sim 25$ s and had a $25-100$ keV fluence 
of $\sim 2.2\times 10^{-5}\,\,{\rm ergs}\,\,{\rm cm}^{-2}$. The redshift of 
the source was determined at $z=2.0369\pm 0.0007$ (Castro et al. 2000), 
yielding a luminosity distance of $D_L=16.9$ Gpc. The optical light curve of 
the afterglow from GRB 000926 fell off as $\sim t^{-1.1\pm 0.2}$ 
for the first $t_{\rm br}=2.0\pm 0.4$ days (break time) and 
subsequently steepened to $\sim t^{-3.2\pm 0.4}$ (Rol, Vreeswijk \& Tanvir 
2000). Furthermore, the resultant sharp break extended over a factor of $\sim 3$ 
in time, inferred from the fitted function of Rol et al. (2000). In addition, 
from the optical to X-ray observations, Sagar et al.  (2000c) have found 
that the spectral index $\bar{\beta}=-0.82\pm 0.02$ at $t=2.26$. Recently, 
Price et al. (2001) also fitted a late-time light curve for this afterglow, 
which slightly flattens as compared with Rol et al.'s (2000). 

In our spreading jet model (with dust extinction),  the spectral index 
$\bar{\beta}=-0.82\pm 0.02$ requires $p=2.6\pm 0.08$, implying $\alpha_1=1.2
\pm 0.06$ (in the slow cooling regime) prior to the break time $t_{\rm br}$
and subsequently $\alpha_2=3.6\pm 0.08$ (from equation [7]). These values are 
in excellent agreement with the observations. Figure 3 shows the observed 
R-band data for GRB 000926 and the theoretically calculated light curve 
based on the model described in section 4. It can be seen from this figure 
that our model provides a good fit to the optical afterglow data. 
More importantly, our model accounts successfully for the observed 
sharpness of the break that appears in the R-band light curve. We note that 
the previous spreading jet model (without dust extinction) seems inconsistent 
with the observed data. This is both because in the previous jet model 
the temporal decay index increases from $\sim 1.3$ to $\sim 2.6$ if $p=2.6$ 
and because the sharpness of the observed break cannot be produced 
in that model, as shown in the present paper and in other studies 
(e.g., Moderski et al. 2000; Kumar \& Panaitescu 2000; Huang et al. 2000a, b, c).  
 
\section{Discussion and Conclusions}

We have discussed the afterglow emission when a highly collimated jet is 
spreading laterally and expanding radially in a circumstellar cloud, and have 
analytically found that the flux decay of the observational optical afterglow 
changes from $\propto t^{-3(p-1)/4}$ (or $t^{-(3p-2)/4}$) in the relativistic 
spherical-like phase to $\propto t^{-(p+1)}$ in the relativistic sideways-expansion 
phase because of 
strong dust extinction outside the initial solid angle of the jet. This provides 
a natural explanation for very rapidly fading afterglows. We have also 
performed numerical calculations based on the generic afterglow model 
proposed by Huang et al. (1999) and developed by Huang et al. (2000c). 
Our numerical results show that one break in the light curve of the optical 
afterglow extends over a factor of $\sim 3$ in time in our present model rather 
than one decade in time in the previous jet model. These results may provide 
a signature for GRBs in dense clouds. In addition, we have given 
a good fit to the observed R-band data of the afterglow from GRB 000926.   

It is interesting to discuss two effects of dust on afterglows. 
First, thermal reradiation and scattering outside the destruction radius 
in the circumstellar giant cloud are expected to cause delayed IR emission 
(Waxman \& Draine 2000; Esin \& Blandford 2000), which has been proposed 
to account for the late-time afterglow feature of GRB 980326 and GRB 970228 
instead of the supernova-like component explanation. At $z=2$ (e.g., GRB 000926), 
the time delay and the flux for the emission at $2.2\,\,\mu$m can be estimated 
by $t_{\rm IR}\sim 1.5\times 10^{7}(\theta_0/0.1)^2(L_{1-7.5}/10^{50}{\rm ergs}\,\,
{\rm s}^{-1})^{1/2}$ s, and $F_{2.2\,\mu{\rm m}}\sim 0.2(L_{1-7.5}
/10^{50}{\rm ergs}\,\,{\rm s}^{-1})^{1/2}\,\,\mu$Jy, respectively. This IR flux 
is expected to be detected in about six months after the burst. Second, since the 
visual extinction is $\sim 50$, the X-rays could be both scattered by the dust 
grains (e.g., M\'esz\'aros \& Gruzinov [2000]),  whose scattering optical depth 
at the X-ray energy $\epsilon_\gamma$ is $\tau\approx 15(\epsilon_\gamma/1
\,{\rm keV})^{-2}$, and efficiently heat the grains. Also, the soft X-rays could 
be absorbed in the giant dense cloud because of a high optical depth
for photo-ionization. These effects may lead to the rapidly fading X-ray 
afterglow of GRB 000926.      

\section*{Acknowledgments}
 
We wish to thank the referee, Dr. Ralph Wijers, for valuable comments, 
and X. Y. Wang for useful discussions. This work was supported 
partially by the National Natural Science Foundation of China (grants 19825109, 
10003001 and 19773007), by the National 973 Project (NKBRSF G19990754), and 
by the Special Funds for Major State Basic Research Projects.

\newpage
\begin{figure}
\centerline{\hbox{\psfig{figure=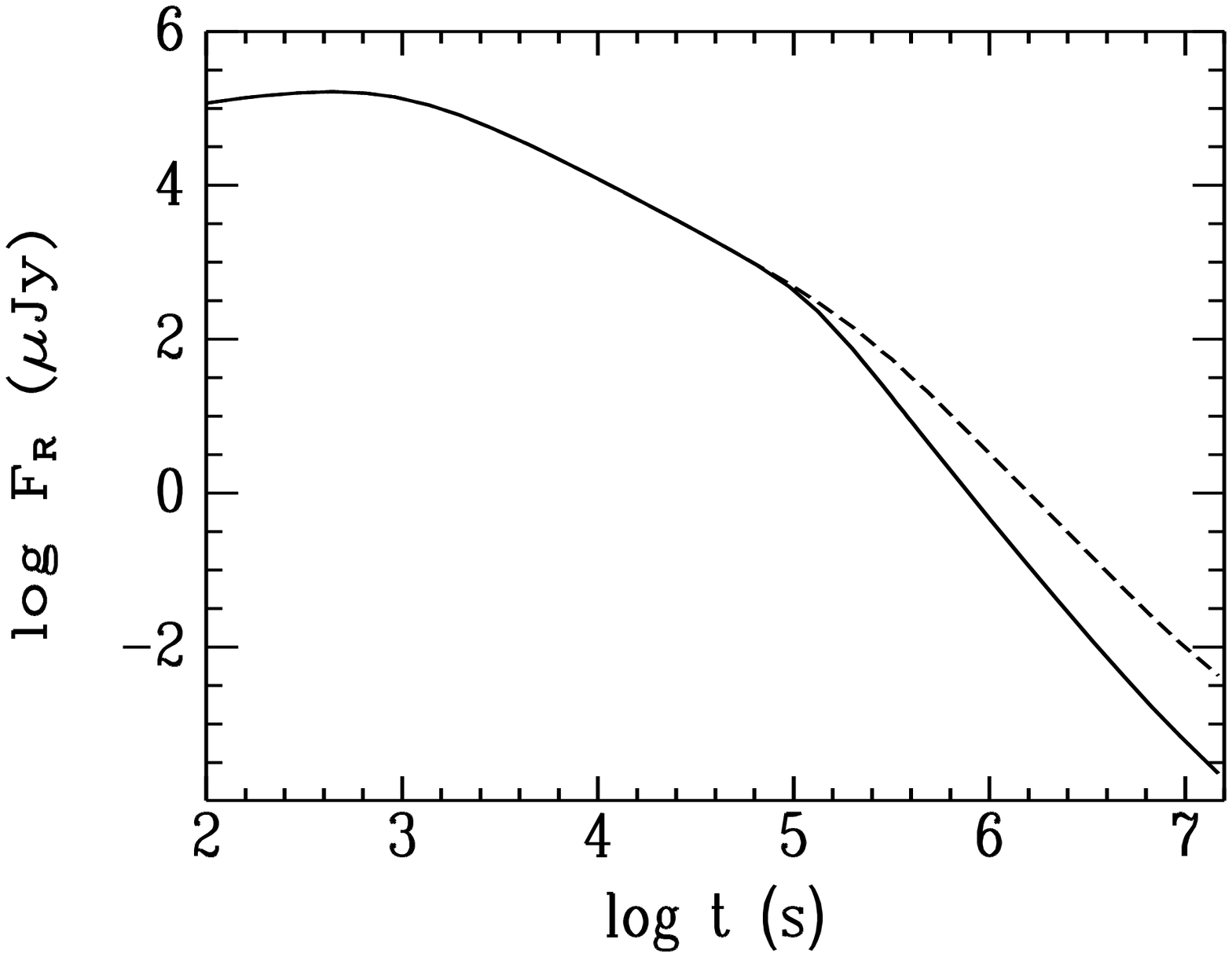,width=7.0in,angle=270}}}
\caption {R-band light curves for dust extinction ({\em solid line}) and no dust extinction 
({\em dashed line}). The model parameters are chosen: $\varepsilon=
10^{54}\,{\rm ergs}\,\,{\rm sr}^{-1}$, $\eta=300$, $\theta_0=0.15$ rad, 
$p=2.5$, $n=10^3\,\,{\rm cm}^{-3}$, $\epsilon_e=0.1$, $\epsilon_B=10^{-6}$,
and $D_L=10$ Gpc.}
\end{figure}

\newpage
\begin{figure}
\centerline{\hbox{\psfig{figure=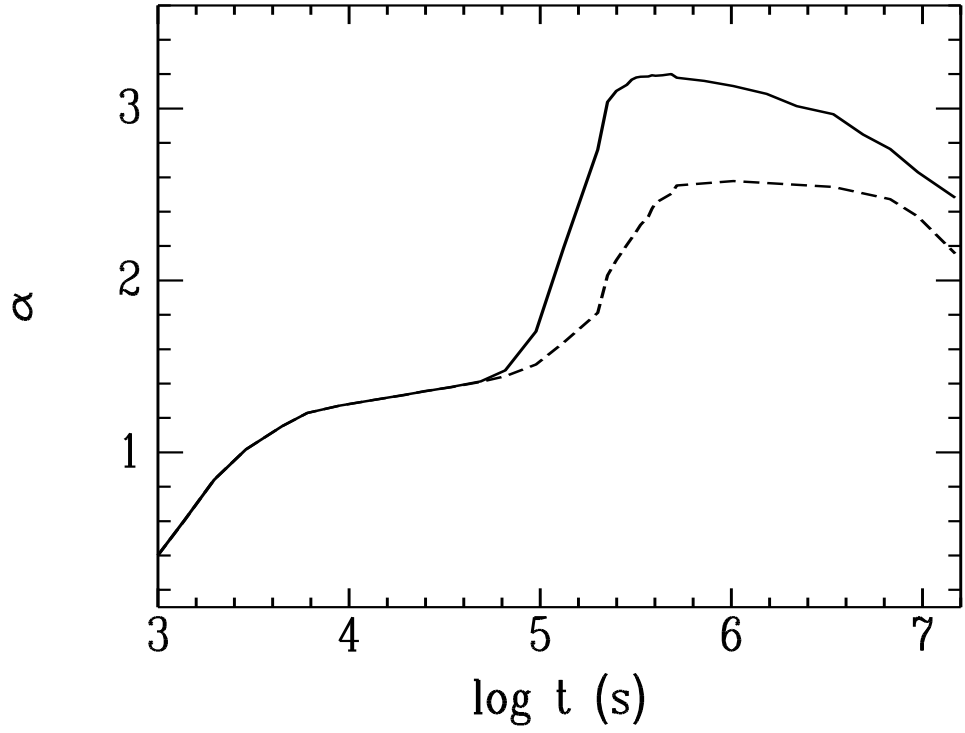,width=7.0in,angle=270}}}
\caption {The R-band $\alpha=-d\ln F_\nu/d\ln t$ as
 a function of time in the cases with 
dust extinction ({\em solid line}) and without dust extinction ({\em dashed line})
for the same parameters as in Figure 1.}
\end{figure}

\newpage
\begin{figure}
\centerline{\hbox{\psfig{figure=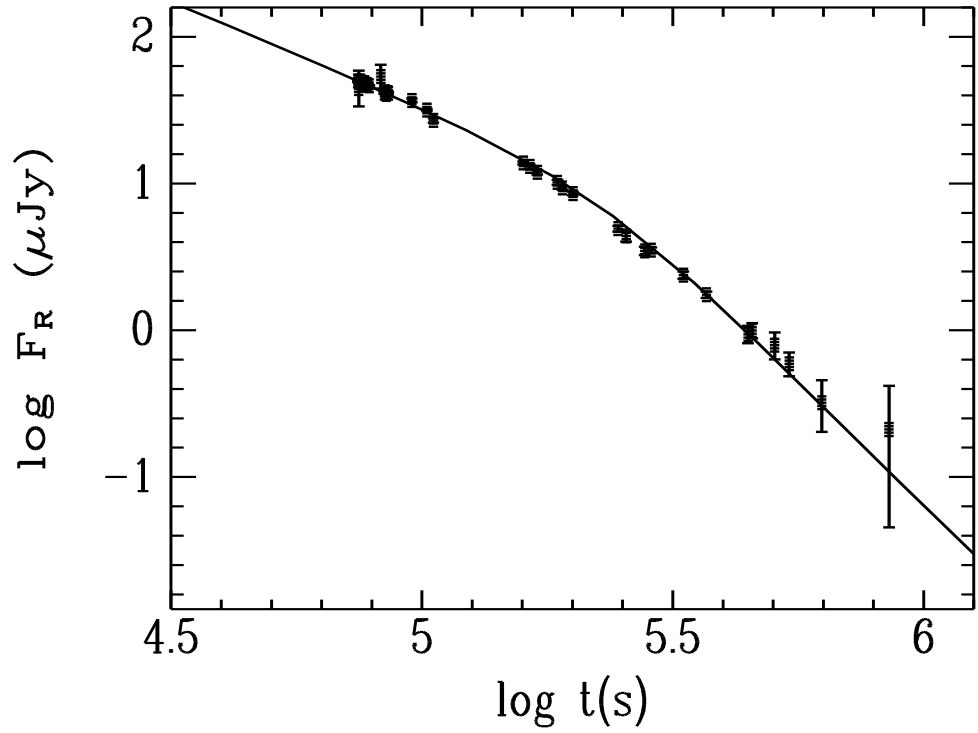,width=7.0in,angle=270}}}
\caption {Comparison between the observed and theoretically calculated light curves 
for the R-band afterglow of GRB 000926. The observational data are taken 
from GCN Circulars (Gorosabel et al. 2000; Dall et al. 2000; Hjorth et al. 
2000a, b; Price et al. 2000; Vrba et al. 2000; Fynbo et al. 2000a, b, c; 
Halpern et al. 2000b, c; Veillet 2000; Rol et al. 2000), and the model light 
curve is calculated for a spreading jet expanding in a circumstellar 
cloud when an observer is located on the jet axis. The model parameters 
are taken: $\varepsilon=8\times10^{53}\,{\rm ergs}\,\,{\rm sr}^{-1}$, 
$\eta=300$, $\theta_0=0.2$ rad, $p=2.6$, $n=10^3\,\,{\rm cm}^{-3}$, 
$\epsilon_e=0.075$, $\epsilon_B=10^{-7}$, and $D_L=16.9$ Gpc.}
\end{figure}
 
\end{document}